\documentstyle[prb,aps,twoside,manuscript,epsf]{revtex}
\begin{document}

\title{\boldmath
Approximate {\it ab initio} calculation of vibrational properties
of hydrogenated amorphous silicon with inner voids
}
\author{
 Serge M. Nakhmanson and
 D. A. Drabold\\
 {\small\it
  Department of Physics and Astronomy, Ohio University,
  Athens, Ohio 45701-2979\\ 
 }
}
\date{\today}
\maketitle

\begin{abstract}
\noindent
We have performed an approximate {\it ab initio} calculation of vibrational
properties of hydrogenated amorphous silicon ({\it a-Si:H})
using a molecular dynamics method. A 216 atom model for pure
amorphous silicon ({\it a-Si})
has been employed as a starting point for our
{\it a-Si:H} models with voids that were made by removing a
cluster of silicon atoms out of the bulk and terminating
the resulting dangling bonds with hydrogens.
Our calculation shows that the presence of voids leads
to localized low energy ($30-50 cm^{-1}$)
states in the vibrational spectrum of the system. The
nature and localization properties of these states are
analyzed by various visualization techniques.

\end{abstract}

\section {Introduction}
It is well known now that typical
{\it a-Si:H}, grown using conventional techniques, 
has certain
characteristic features: it contains a
significant amount of 
nanovoids\cite{Street,Antonio,Postol,Craven}, and
it has an average hydrogen concentration of approximately 10 atomic
percent\cite{Street}.
Hydrogen, present in the bulk of the material, can generally
be found in three different states: it can be bonded, passivating  
dangling silicon bonds in the network or on the surfaces
of the voids, it can be nonbonded (most of it, because the
concentration of dangling bonds in {\it a-Si} is much lower
than concentration of hydrogen atoms\cite{Street}) and it can
also exist in gaseous phase inside the voids.

Important recent work with a new hot wire
form of a-Si:H has recently been reported\cite{NREL}
which might connect a low density of voids with
unique vibrational properties. We have therefore performed 
studies of a dense continuous random network, and models
with hydrogenated voids intentionally introduced to compare
the vibrational modes in the two cases in detail. This study
leads us to believe that hydrogenated voids lead to low
frequency modes, though not at energies small enough to 
be tunneling modes\cite{pafdad}. We are also interested in
experimentally discernible vibrational signatures linked
to voids and or hydrogen.

In Sec.\ II of this Brief Report we present the detailed algorithm
of construction of the models and calculation techniques
used to study their vibrational behavior. Then in Sec.\ III
we discuss our results received for different models of 
{\it a-Si:H} with voids and compare them with the results
obtained for pure {\it a-Si} model.
Finally, in Sec.\ IV we draw some conclusions regarding
the question: do voids lead to low energy states
in {\it a-Si:H}?

\section {Model construction and calculations}
In this section we present our scheme for
constructing a model of hydrogenated amorphous silicon
with a void.
We use a model generated from a 
Wooten, Winer, Weaire\cite{WWW} scheme, involving 216 atoms
for pure {\it a-Si} by Djordjevic, Thorpe and
Wooten\cite{WWWupd} as a ``base'' for construction of
models for {\it a-Si:H} with voids in the bulk.
We label this the DTW model in what follows.
For our MD simulations we employ a code developed by Sankey
and coworkers\cite{SN}, based upon density
functional theory within the local density approximation 
(DFT-LDA)\cite{HK}.
The main features of the method are a spin averaged, non-self-consistent
version of density functional theory based on a linearization of
the Kohn-Sham equations\cite{Harris}, 
a minimal pseudoatomic orbital basis set with an
interaction confinement radius, and nonlocal, norm-conserving
pseudopotentials\cite{ppot}. 
In equilibrium geometry search, we consider
our models fully relaxed when total forces on any atom in the
supercell are less than $0.03$ eV/\AA.

In the first step (which we
need to perform only once) we relax the DTW model, 
obtaining  its equilibrium geometry configuration by 
dynamical quenching. In our case relaxing
the model
resulted only in minor
network rearrangements\cite{JJ}.
The radial distribution function, obtained for the relaxed
model, appeared to be
in good agreement with experimental data, as was the
case for the original model before relaxation.

In the second step we remove a cluster of {\it Si} atoms,
thus creating a void in the silicon network. 
As the removal of some silicon atoms out of the
network results in dangling bonds,
we must terminate these bonds
with hydrogen atoms. In our models, presented in this 
publication, we consider entirely passivated networks,
where {\it every} dangling bond in the network is
terminated by hydrogen atom,
initially placed directly on the bond $1.5$ \AA ~away 
from the {\it Si} atom, that we keep.

After
discarding the chosen cluster of {\it Si} atoms and saturating
the dangling bonds on the surface of the void with hydrogens
we perform the third step --- another MD relaxation,
which gives us the equilibrium configuration for the new
structure with void. At this point the actual number of relaxation
steps
required to make forces on every atom sufficiently small
varies greatly, depending on the size and form of the void.

Once the equilibrium state is reached, we
can proceed with the dynamical matrix calculation,
displacing every atom in the supercell in three orthogonal
directions (by $0.03$~\AA, which is suitable for
this purpose) and computing the resulting spring constants 
as second derivatives
of the total energy of the system.
Diagonalizing the dynamical matrix we finally receive its
eigenvalues together with the corresponding eigenvectors,
which enables us to carry out the full investigation of
vibrational behavior of a given model.

We plot  a Vibrational Density of States (VDOS) 
for a set of dynamical matrix eigenvalues
$E_i, i = 1 \ldots N$,  where
$N$ is the number of atoms in the system,
which gives us information about vibrational states distribution 
along the energy axis.
In the figures, we use a gaussian broadened
form for $\delta(E-E_i)$, with width 10 $cm^{-1}$.
Then, in order to understand the localization properties of
vibrational states, we construct Inverse Participation
Ratio\cite{ipr} (IPR) graphs.

Finally, for any energy mode we can estimate vibrational
activity for every separate atom, examining the components
of the eigenvector ``belonging'' to this particular atom and
comparing, for example, the sum of their squares 
on the reference atom, with the
sum over all atoms in the supercell. After forming this
chart of individual atomic IPRs we can employ
this information for dynamical animation of vibrational mode
(by creating a file, consisting of set of frames, each
containing coordinates and displacement vector 
components for every
atom in the supercell, which can be used as an input for
{\it Xmol} molecular display and animation program\cite{Xmol}) or
for creating its static equivalent in a form of a grey scale
(where atoms are assigned different shades according to
their vibrational activity).

\section {Discussion of results}

In this section we will discuss our results received for
two different {\it a-Si:H} models with voids. The first model
has five {\it Si} atoms removed and twelve hydrogens added
(we refer to it as ``small bubble'' model),
the second one has 23 {\it Si} atoms removed and 36 hydrogens
added (``big bubble'' model).

To be able to compare the results for our
{\it a-Si:H} models and results, obtained
for pure {\it a-Si} DTW model, which serves as our
``reference'' model, we have performed the set of all
calculations, described in the previous section, for DTW
as well as for both models containing voids.

\begin{figure}

\epsfxsize=10cm
\moveright 3cm \vbox{\epsfbox{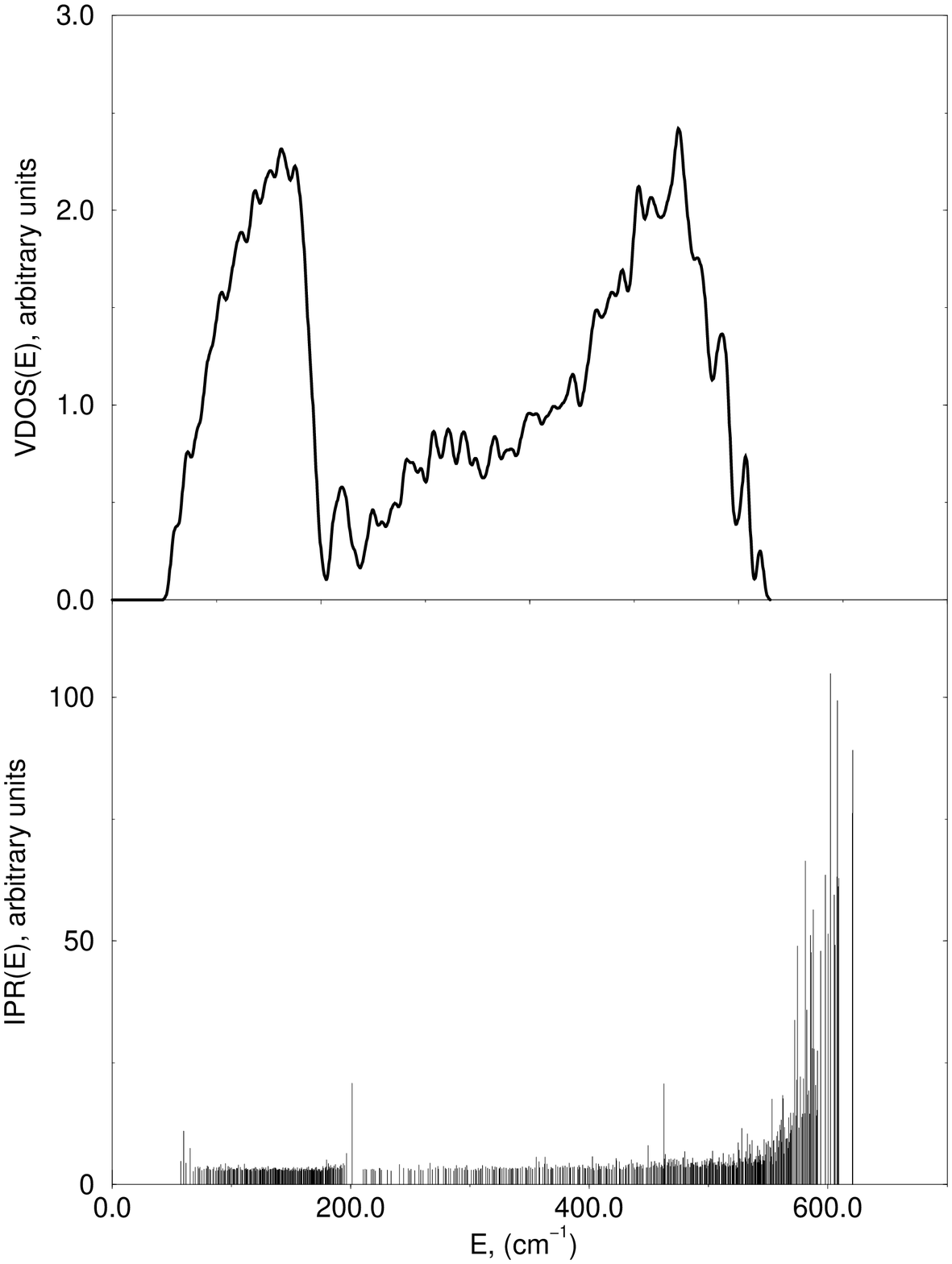}}

\vskip 2cm

\caption{
\label{pure_si}
Vibrational density of states (upper panel)
and inverse participation ratio (lower panel)
for 216 atom DTW model for pure {\it a-Si}.}
\end{figure}

\begin{figure}

\epsfxsize=10cm
\moveright 3cm \vbox{\epsfbox{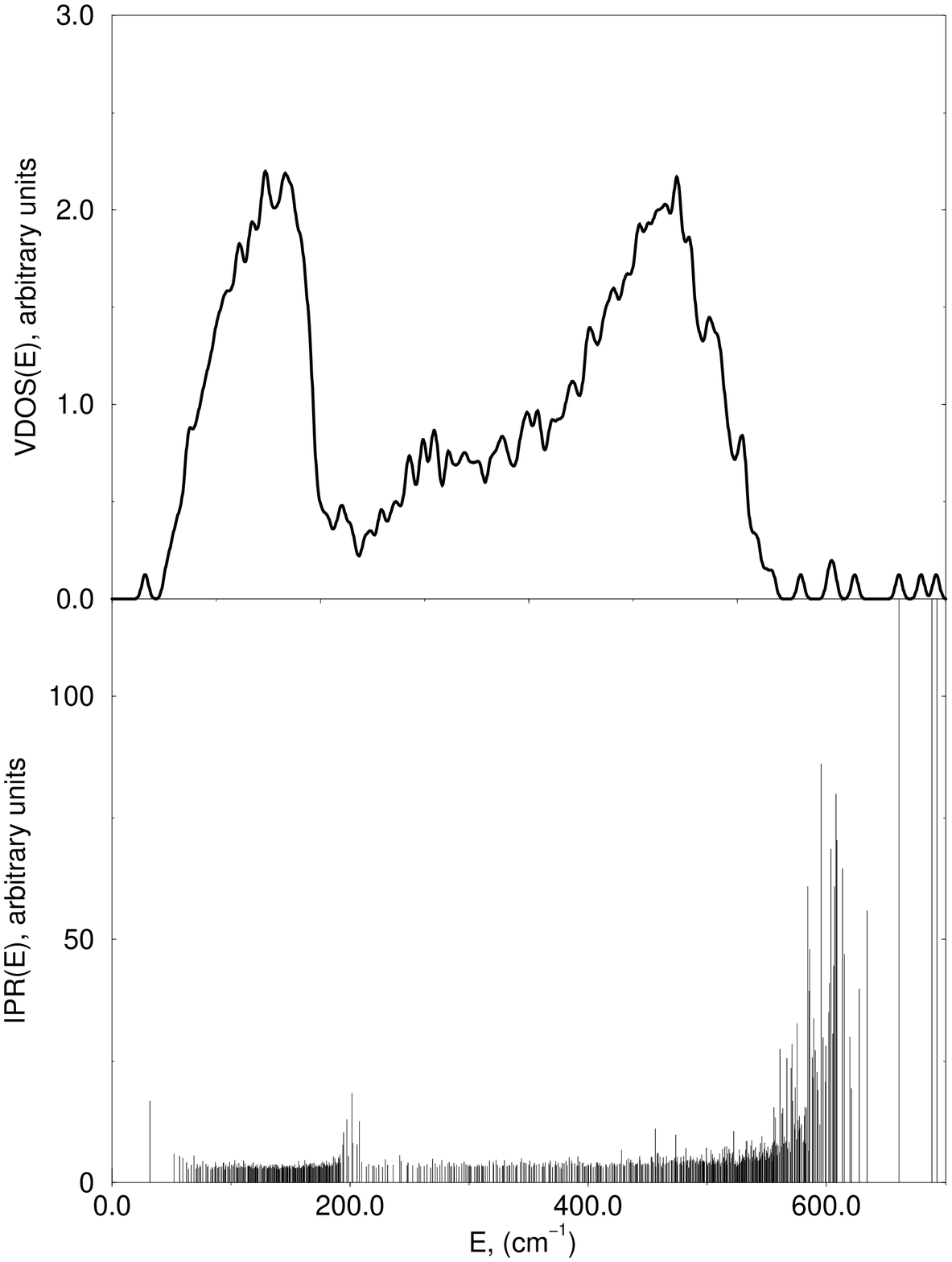}}

\vskip 2cm

\caption{
\label{s_bubble}
Vibrational density of states
and inverse participation ratio 
for ``small bubble'' model (211 {\it Si}
atoms, 12 {\it H} atoms) for {\it a-Si:H}
with void (high energy modes are not shown).}
\end{figure}

\begin{figure}

\epsfxsize=10cm
\moveright 3cm \vbox{\epsfbox{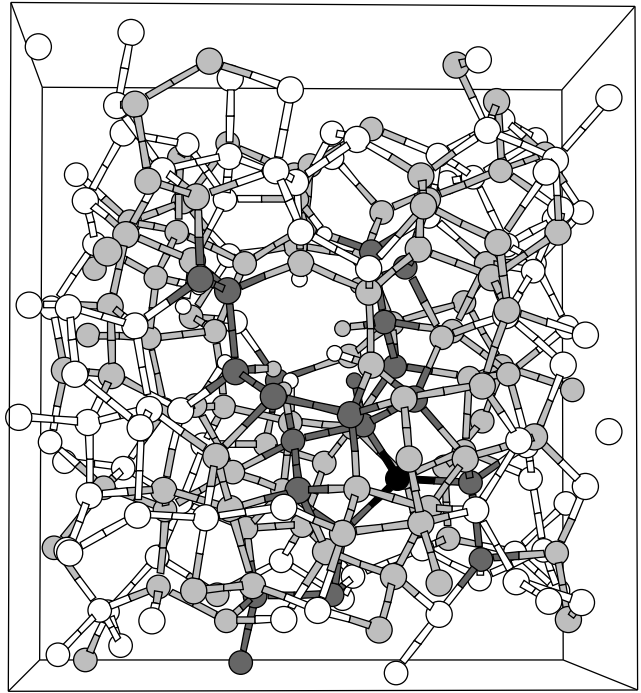}}

\caption{
\label{sb_map}
Localized low energy vibrational mode
for ``small bubble'' model. Vibrational
energy is 32.0~$cm^{-1}$.
Here every atom pictured in black accounts for
more than 10\% of total supercell excitation.
Atoms, represented by dark grey, light grey
and white, account for more than 1 and up to 10\%, 
more than 0.1 and up to 1\% and
less than or equal to 0.1 \% of total supercell
excitation accordingly.}
\end{figure}

\begin{figure}

\epsfxsize=10cm
\moveright 3cm \vbox{\epsfbox{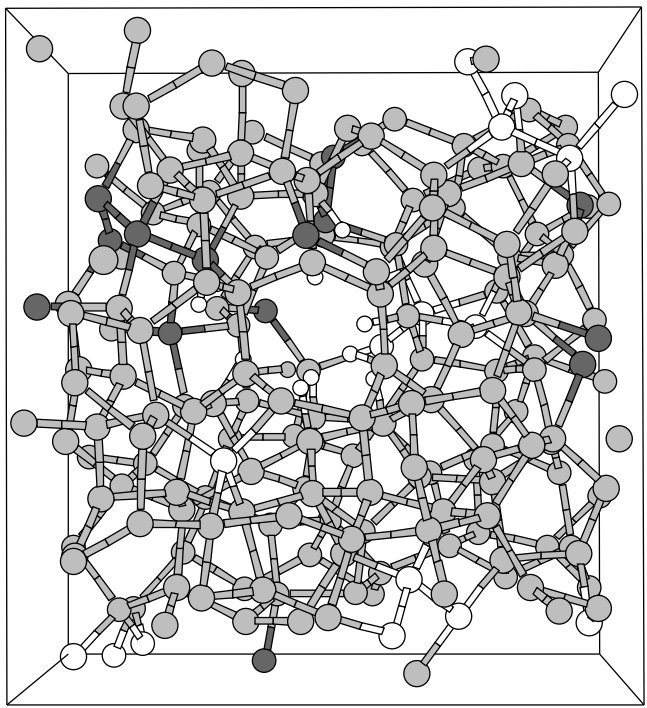}}

\caption{
\label{sb_map_deloc}
Low energy vibrational mode
for ``small bubble'' model which is
delocalized. Vibrational
energy is 64.4~$cm^{-1}$.
Grey scale conventions the same as for
FIG.\ \ref{sb_map}. }
\end{figure}

In Fig.\ \ref{pure_si} our results for pure {\it a-Si}
model are presented; note that the calculations show
that there are no vibrational states present with 
energies up to approximately 55--60~$cm^{-1}$.
Now, if we compare these results with the results
obtained for the ``small bubble'' model, shown on
Fig.\ \ref{s_bubble}, and restrict our attention only
to low energy states, we can see that for a model with void
a new vibrational mode emerges at 32~$cm^{-1}$ (which is
practically right in the middle of the low energy gap in
{\it a-Si} vibrational spectrum)
and this mode has high IPR and can be considered
spatially localized. The grey scale map
for the aforementioned mode is presented in Fig.\ \ref{sb_map}.
It enables us to estimate where this vibration is
localized in the supercell and how it decays in space. We can
see that the mode has rather complicated structure, although
it is mostly localized around the surface of the
void and decays very rapidly when we move away from the void,
there are certain directions where it decays more slowly and
a whole cluster of vibrationally active atoms to the side of 
the void. For comparison in
Fig.\ \ref{sb_map_deloc} we present one of the low energy
(64.4~$cm^{-1}$) modes for ``small bubble'' model that has
relatively low IPR and, according to the picture, is rather
uniformly distributed in space.

\begin{figure}

\epsfxsize=10cm
\moveright 3cm \vbox{\epsfbox{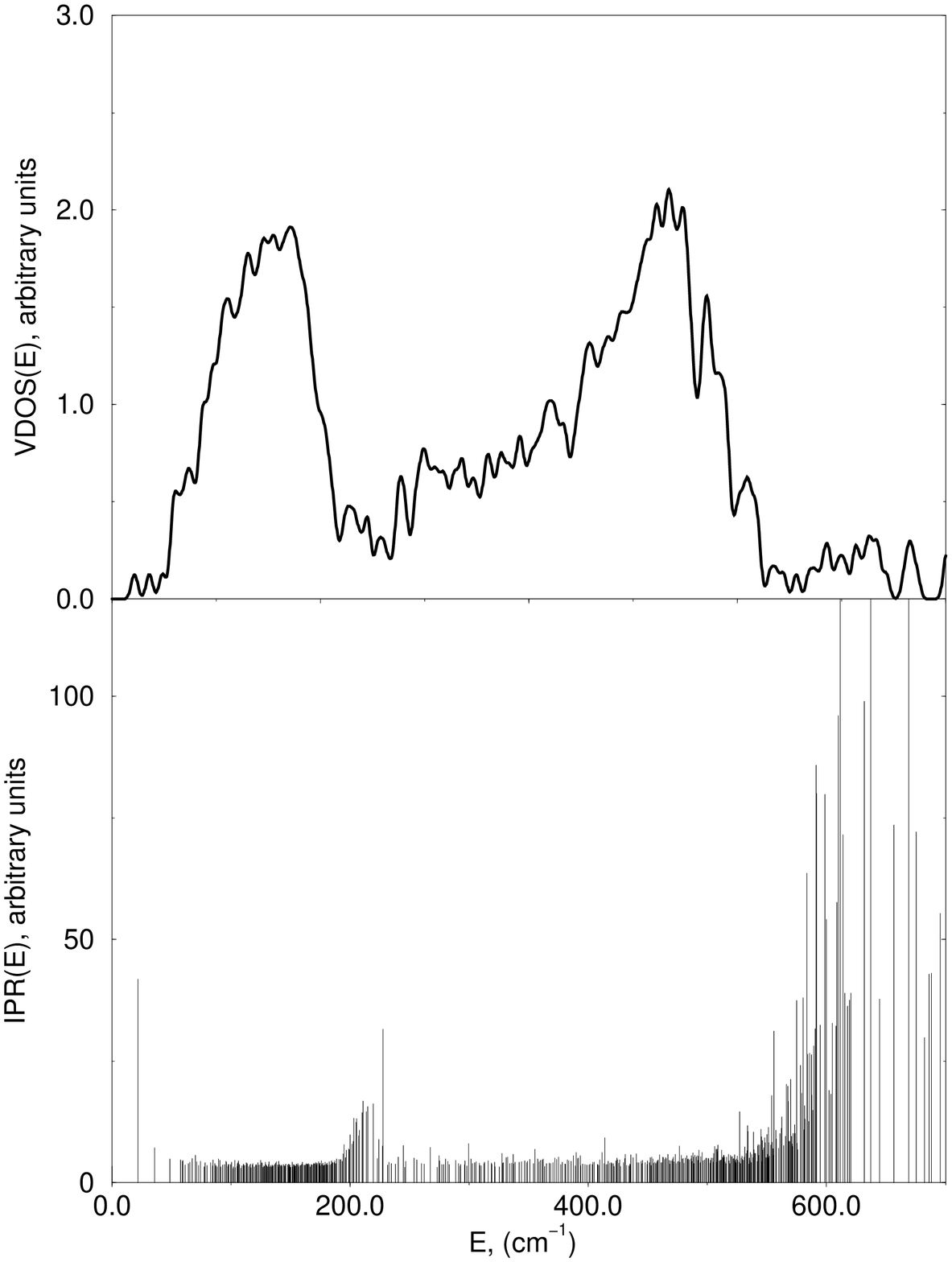}}

\vskip 2cm

\caption{
\label{b_bubble}
Vibrational density of states
and inverse participation ratio 
for ``big bubble'' model (193 {\it Si}
atoms, 36 {\it H} atoms) for {\it a-Si:H}
with void (high energy modes are not shown).}
\end{figure}

\begin{figure}

\epsfxsize=10cm
\moveright 3cm \vbox{\epsfbox{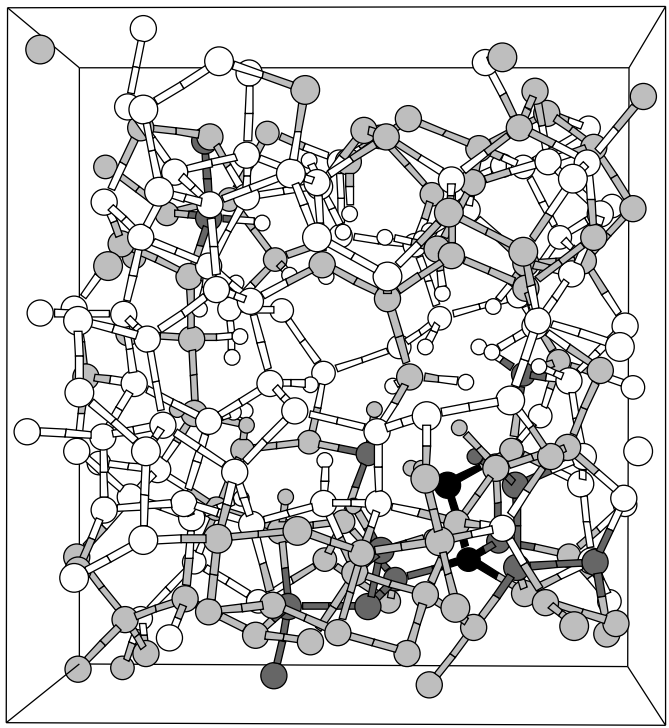}}

\caption{
\label{bb_map}
Localized low energy vibrational mode
for ``big bubble'' model. Vibrational
energy is 21.8~$cm^{-1}$.
Grey-scale conventions the same as for
FIG.\ \ref{sb_map}.}
\end{figure}

Now comparing the results for the first two models, discussed
in the previous paragraph, with the results received for the
``big bubble'' model (Fig.\ \ref{b_bubble})
we can see that now we have three new low energy states, but
only one of them, at 21.8~$cm^{-1}$ has high IPR (more than
two times larger than for the similar mode in ``small bubble''
model) or is localized. Examining the grey scale map for this mode,
presented in Fig.\ \ref{bb_map} we also notice that spatial
localization of the mode is quite similar to ``small bubble''
model mode with the exception that the former decays much faster
and its localization on a cluster of {\it Si} atoms to the
side of the void is much sharper.

Our hydrogenated models also produce highly localized states
at 600--630 $cm^{-1}$ (hydrogen bend) and
2000--2200 $cm^{-1}$ (hydrogen stretch). Other hydrogen
related states, corresponding to mixing of the first two,
fill out the region of 700--1000 $cm^{-1}$.

\section {Conclusions}
Studying the models for {\it a-Si:H} we
have found localized low energy 
modes in vibrational spectrum of the system.
The nature of these modes is quite complicated 
but evidently connected with the existence of voids ---
for both localized low energy modes considered a 
number of silicon atoms located close to the surface of the 
void exhibits high vibrational activity. 
We can call the low energy modes 
``floppy'' because  
the introduction of 
void-type defects into a
silicon network 
significantly reduces the coordination numbers for atoms
situated around the defect (thus making this particular
region of the network ``floppy''), but, due to their
local origin, these modes are not exactly the same as
the celebrated Phillips-Thorpe ``floppy'' modes, associated with
the {\it average} coordination in the system.

The principal shortcoming of this calculation is its
finite size, which can be interpreted as an artificial 
void-void interaction, or a very high density of the voids.
For the ``small bubble" model, this is apparently not
a serious problem, since the void-induced low energy state
is well localized. The main point of this work is that
voids lead to low frequency modes, and contribute (along
with extended low frequency modes) to the vibrational
spectrum. Such modes could contribute to thermal
conductivity by resonant mixing with each other and
acoustic modes\cite{JJ}.

Also, in connection with the important discoveries of 
Ref. \cite{NREL}, this work suggests that indeed the
small density of voids in their material should lead
to a reduction in low energy (but not tunneling
mode energy) vibrational excitations. It is quite possible
that the tunneling modes are not directly accessible through
the usual harmonic approximation we have invoked here, and
require other methods (such as direct dynamical simulation for
long times) to be studied.

We see that the origin of tunneling state from defects
(not disorder) has been observed in experiments\cite{NREL2}.
Because we see an expected decrease in mode frequency
with increasing void size and the "no-friction" material
of Liu {\it et al.\  } has few voids, we {\it speculate}
that tunneling modes are just very low frequency void 
surface excitations of voids larger than we discuss here.

\section*{Acknowledgments}
We thank Dr.\ M.\ Cobb for providing us a program for
vibrational mode animation. We also acknowledge helpful
conversations with Dr.\ P.\ Allen, Dr.\ R.\ Biswas and
Dr.\ P.\ A.\ Fedders.
This work was supported by NSF
under Grants number DMR 96-18789 and DURIP N00014-97-1-0315.


\end{document}